\documentclass[11pt]{article}
\pdfoutput=1  
\usepackage{graphicx,color}
\usepackage{appendix}
\usepackage{latexsym,amsmath,amssymb,graphicx,booktabs}
\usepackage{epsfig,latexsym,cite}
\usepackage{hyperref}
\numberwithin{equation}{section}

\definecolor{MyBlue}{rgb}{0.15,0.15,0.70}

\hypersetup{
colorlinks=true,
citecolor=MyBlue,
linkcolor=MyBlue,
urlcolor=MyBlue
}

\input epsf
\usepackage{amssymb}
\usepackage{amsmath}
\usepackage{amsfonts}
\usepackage{upgreek}
\usepackage{latexsym}

\newcommand{\iBox}{\Box^{-1}}
\newcommand{\Stu}{St\"uckelberg }

\renewcommand\({\left(}
\renewcommand\){\right)}
\renewcommand\[{\left[}
\renewcommand\]{\right]}

\newcommand\n{{\mbox {\boldmath $\nabla$}}}
\newcommand{\ra}{\rightarrow}

\def\lsim{\raise 0.4ex\hbox{$<$}\kern -0.8em\lower 0.62
ex\hbox{$\sim$}}

\def\gsim{\raise 0.4ex\hbox{$>$}\kern -0.7em\lower 0.62
ex\hbox{$\sim$}}

\def\lbar{{\hbox{$\lambda$}\kern -0.7em\raise 0.6ex
\hbox{$-$}}}

\newcommand\eq[1]{eq.~(\ref{#1})}
\newcommand\eqs[2]{eqs.~(\ref{#1}) and (\ref{#2})}

\newcommand\pa{\partial}
\newcommand\p{\partial}

\newcommand\ee{\end{equation}}
\newcommand\be{\begin{equation}}
\def\bea{\begin{array}}
\def\eea{\end{array}}\def\ea{\end{array}}
\newcommand\ees{\end{eqnarray}}
\newcommand\bees{\begin{eqnarray}}
\def\nn{\nonumber}

\def\a{\alpha}

\def\s{\sigma}

\def\d{\delta}

\def\eps{\epsilon}

\def\dslash{\hspace{-1mm}\not{\hbox{\kern-2pt $\partial$}}}
\def\Dslash{\not{\hbox{\kern-4pt $D$}}}
\def\pslash{\not{\hbox{\kern-2.1pt $p$}}}
\def\kslash{\not{\hbox{\kern-2.3pt $k$}}}
\def\qslash{\not{\hbox{\kern-2.3pt $q$}}}

\newcommand{\vk}{{\bf k}}

\def\p1{{\bf p}_1}
\def\p2{{\bf p}_2}
\def\k1{{\bf k}_1}
\def\k2{{\bf k}_2}

\newcommand{\emn}{\eta_{\mu\nu}}

\newcommand{\eMN}{\eta^{\mu\nu}}
\newcommand{\eRS}{\eta^{\rho\sigma}}
\newcommand{\eMR}{\eta^{\mu\rho}}
\newcommand{\eNS}{\eta^{\nu\sigma}}
\newcommand{\eMS}{\eta^{\mu\sigma}}
\newcommand{\eNR}{\eta^{\nu\rho}}

\newcommand{\gmn}{g_{\mu\nu}}

\newcommand{\gMN}{g^{\mu\nu}}

\newcommand{\hmn}{h_{\mu\nu}}
\newcommand{\hrs}{h_{\rho\sigma}}

\newcommand{\hMN}{h^{\mu\nu}}

\newcommand{\bhmn}{\bar{h}_{\mu\nu}}

\newcommand{\xim}{\xi_{\mu}}
\newcommand{\xin}{\xi_{\nu}}

\newcommand{\pam}{\pa_{\mu}}

\newcommand{\pan}{\pa_{\nu}}
\newcommand{\parho}{\pa_{\rho}}

\newcommand{\paM}{\pa^{\mu}}
\newcommand{\paN}{\pa^{\nu}}
\newcommand{\paR}{\pa^{\rho}}
\newcommand{\paS}{\pa^{\sigma}}

\newcommand{\Gmn}{G_{\mu\nu}}

\newcommand{\Tmn}{T_{\mu\nu}}
\newcommand{\Smn}{S_{\mu\nu}}

\newcommand{\TMN}{T^{\mu\nu}}

\newcommand{\dddM}{\kern 0.2em \raise 1.9ex\hbox{$...$}\kern -1.0em \hbox{$M$}}
\newcommand{\dddQ}{\kern 0.2em \raise 1.9ex\hbox{$...$}\kern -1.0em \hbox{$Q$}}
\newcommand{\dddI}{\kern 0.2em \raise 1.9ex\hbox{$...$}\kern -1.0em\hbox{$I$}}
\newcommand{\dddJ}{\kern 0.2em \raise 1.9ex\hbox{$...$}\kern-1.0em
\hbox{$J$}}
\newcommand{\dddcalJ}{\kern 0.2em \raise 1.9ex\hbox{$...$}\kern-1.0em
\hbox{${\cal J}$}}

\newcommand{\dddO}{\kern 0.2em \raise 1.9ex\hbox{$...$}\kern -1.0em
\hbox{${\cal O}$}}
\def\dddz{\raise 1.5ex\hbox{$...$}\kern -0.8em \hbox{$z$}}
\def\dddd{\raise 1.8ex\hbox{$...$}\kern -0.8em \hbox{$d$}}
\def\dddbd{\raise 1.8ex\hbox{$...$}\kern -0.8em \hbox{${\bf d}$}}
\def\ddbd{\raise 1.8ex\hbox{$..$}\kern -0.8em \hbox{${\bf d}$}}
\def\dddx{\raise 1.6ex\hbox{$...$}\kern -0.8em \hbox{$x$}}

\newcommand{\ode}{\Omega_{\rm DE}}

\begin{document}

\begin{titlepage}

\vspace*{2cm}

\centerline{\Large \bf Apparent ghosts and spurious degrees}

\vspace{5mm}

\centerline{\Large \bf of freedom in non-local theories}

\vskip 0.4cm
\vskip 0.7cm
\centerline{\large Stefano Foffa, Michele Maggiore and Ermis Mitsou}
\vskip 0.3cm
\centerline{\em D\'epartement de Physique Th\'eorique and Center for Astroparticle Physics,}  
\centerline{\em Universit\'e de Gen\`eve, 24 quai Ansermet, CH--1211 Gen\`eve 4, Switzerland}

\vskip 1.9cm

\begin{abstract}

Recent work has shown that non-local modifications of the Einstein equations can have interesting cosmological consequences and can provide a dynamical origin for dark energy, consistent with existing data.
At first sight these theories are plagued by ghosts. We show that
these apparent ghost-like instabilities do not describe actual propagating degrees of freedom, and there is no issue of ghost-induced quantum vacuum decay.

\end{abstract}


\end{titlepage}

\newpage

\section{Introduction}

Non-local classical equations and non-local field theories have been sporadically 
studied since the early days of field theory~\cite{Pais:1950za}. In general, they present a number of subtle issues concerning ghosts and unitarity, and often it is not obvious even  how many degrees of freedom they describe, see e.g. 
\cite{Eliezer:1989cr,Simon:1990ic,Woodard:2006nt,Barnaby:2007ve,Ilhan:2013xe}. In recent years there has been a renewed interest for non-local models in connection with cosmology.
Non-local effective equations of motion can play an important role in 
explaining the ``old" cosmological constant problem, through a degravitation mechanism that promotes Newton's constant to a non-local operator \cite{ArkaniHamed:2002fu,Dvali:2007kt} (see also~\cite{Dvali:2000xg,Dvali:2002pe,Barvinsky:2003kg,Dvali:2006su}), and non-local cosmological models have interesting observational consequences and can play a role in explaining the origin of dark energy
\cite{Deser:2007jk,Koivisto:2008xfa,Koivisto:2008dh,Capozziello:2008gu,Barvinsky:2011hd,Barvinsky:2011rk,Elizalde:2011su,Zhang:2011uv,Elizalde:2012ja,Park:2012cp,Bamba:2012ky,Deser:2013uya,Ferreira:2013tqn,Dodelson:2013sma}. 
 Non-local gravity models have also been studied as UV modifications of GR, see e.g.
\cite{Hamber:2005dw,Khoury:2006fg,Biswas:2010zk,Modesto:2011kw,Briscese:2012ys}.

In two recent papers \cite{Jaccard:2013gla,Maggiore:2013mea} we have proposed a non-local approach that allows us to introduce a mass term in the Einstein equations, 
in such a way that   the invariance under diffeomorphisms is not spoiled, and we do not need to introduce an external reference metric (contrary to what happens in the conventional local approach to massive gravity
\cite{deRham:2010ik,deRham:2010kj,deRham:2011rn,Hassan:2011hr,Hassan:2011tf,Hassan:2011ea,Hinterbichler:2011tt}). In particular, in \cite{Maggiore:2013mea} has been proposed a classical model based on the non-local equation
\be\label{modela2}
\Gmn -\frac{d-1}{2d}\, m^2\(\gmn \iBox_{\rm ret} R\)^{\rm T}=8\pi G\,\Tmn\, .
\ee
Here  $d$ is the number of spatial dimensions  (the factor $(d-1)/2d$ is a convenient normalization of the parameter $m^2$),  $\Box=\gMN\n_{\mu}\n_{\nu}$ is the d'Alembertian operator with respect to the metric $\gmn$ and, quite crucially,
$\iBox_{\rm ret}$ is its inverse computed with the retarded Green's function.
The superscript T denotes the extraction of the transverse part of the tensor, which exploits the fact that, in a generic Riemannian manifold, any symmetric tensor $\Smn$ can be decomposed as
$S_{\mu\nu}=
S_{\mu\nu}^{\rm T}+\frac{1}{2}(\n_{\mu}S_{\nu}+\n_{\nu}S_{\mu})$, with
$\n^{\mu}S_{\mu\nu}^{\rm T}=0$ \cite{Deser:1967zzb,York:1974}. The extraction of the transverse part of a tensor is itself a non-local operation, which involves further $\iBox$ operators. 
For instance in flat space, where $\n_{\mu}\ra\pam$, it is easy to show that
\be\label{Tflat}
S_{\mu\nu}^{\rm T}=\Smn
-\frac{1}{\Box}(\pam\paR S_{\rho\nu}+\pan\paR S_{\rho\mu})
+\frac{1}{\Box^2}\pam\pan\paR\paS S_{\rho\sigma}\, .
\ee
Again, in \eq{modela2}
all  $\iBox$ factors coming from the extraction of the transverse part are defined with the retarded Green's function, so that \eq{modela2} satisfies causality. Furthermore, since the left-hand side of \eq{modela2} is transverse by construction, the energy-momentum tensor is automatically conserved, $\n^{\mu}\Tmn=0$. Both  causality and energy-momentum conservation were lost in the original degravitation proposal \cite{ArkaniHamed:2002fu}, and in this sense \eq{modela2} can be seen as a refinement of the original idea. However, the explicit appearance of retarded Green's function in the equations of motion has important consequences for the conceptual meaning of 
an equation such as (\ref{modela2}), as we will discuss below.

As shown in \cite{Maggiore:2013mea,FMM}, \eq{modela2} has very interesting cosmological properties, and in particular it generates a dynamical dark energy. Since during radiation dominance (RD) the Ricci scalar $R$ vanishes, the term $\iBox R$ starts to grow only during matter dominance (MD), thereby providing in a natural way a delayed onset of the accelerated expansion (similarly to what happens in the model proposed in 
\cite{Deser:2007jk}). Furthermore, this model is highly predictive since it only introduces a single parameter $m$, that replaces the cosmological constant in $\Lambda$CDM. In contrast,  models based on quintessence, $f(R)$-gravity, or the non-local model of \cite{Deser:2007jk} in which a term $Rf(\iBox R)$ is added to the Einstein action, all introduce at least one arbitrary function, which is typically tuned so to get the desired cosmological behavior. In our case, we can fix the value of $m$ so to reproduce the observed value $\ode\simeq 0.68$.
This gives  $m\simeq 0.67 H_0$, and  leaves us with no free parameter.  We then get a pure prediction for the EOS parameter of dark energy. Quite remarkably, writing
$w_{\rm DE}(a)=w_0+(1-a) w_a$,  in  \cite{Maggiore:2013mea} we found
$w_0\simeq-1.04$ and $w_a\simeq -0.02$, consistent with the Planck data, and on the phantom side.

These cosmological features make \eq{modela2} a potentially very attractive dark energy model. The presence of the $\iBox$ operator raises however a number of potential problems of theoretical consistency, and the purpose of this paper is to investigate them in some detail. The crucial problem can  already be seen linearizing \eq{modela2} over flat space. Writing $\gmn =\emn+\kappa\hmn$, where $\kappa= (32\pi G)^{1/2}$ and $\emn=(-,+,\ldots ,+)$, the equation of motion of this theory takes the form
\be\label{line1}
{\cal E}^{\mu\nu,\rho\sigma}\hrs
-\frac{d-1}{d}\, m^2 P_{\rm ret}^{\mu\nu}P_{\rm ret}^{\rho\sigma}
\hrs=-16\pi G\TMN\, ,
\ee
where  ${\cal E}^{\mu\nu,\rho\sigma}$  is the  Lichnerowicz operator, while
\be
P_{\rm ret}^{\mu\nu}=\eMN-\frac{\paM\paN}{\Box_{\rm ret}}\, , 
\ee
and  $1/\Box_{\rm ret}$ is the  retarded inverse of the  flat-space d'Alembertian. 
Apparently, the corresponding quadratic Lagrangian is 
\be\label{Lquadr}
{\cal L}_2=\frac{1}{2}\hmn{\cal E}^{\mu\nu,\rho\sigma}\hrs
-\frac{d-1}{2d}\, m^2\hmn P^{\mu\nu}P^{\rho\sigma}\hrs\, .
\ee
Adding the usual gauge fixing term of linearized massless gravity,
${\cal L}_{\rm gf}=-(\paN\bhmn ) (\parho\bar{h}^{\rho\mu})$, and inverting the resulting quadratic form we get the propagator
\be\label{prop}
\tilde{D}^{\mu\nu \rho\s}(k)=\frac{-i}{2k^2}\, 
\( \eMR\eNS +\eMS\eNR-\frac{2}{d-1}\eMN\eRS \) - \frac{1}{d(d-1)}\, \frac{im^2}{k^2(-k^2+m^2)}\eMN\eRS\, ,
\ee
plus terms proportional to $k^{\mu}k^{\nu}$, $k^{\rho}k^{\sigma}$ and
$k^{\mu}k^{\nu}k^{\rho}k^{\sigma}$, that give zero when contracted with a conserved energy-momentum tensor.
The first term is the usual propagator of a massless graviton, for $d$ generic. The term proportional to $m^2$ gives  an extra contribution to  
the saturated propagator $\tilde{T}_{\mu\nu}(-k)
\tilde{D}^{\mu\nu \rho\s}(k)\tilde{T}_{\rho\sigma}(k)$, equal to
\be\label{TT}
\frac{1}{d(d-1)}\tilde{T}(-k)\[ -\frac{i}{k^2}-\frac{i}{(-k^2+m^2)}\]
\tilde{T}(k)\, .
\ee
This term apparently describes the exchange of a healthy massless scalar plus a ghostlike massive scalar. 
In general, a ghost has two quite distinct effects:  at the classical level, it can give rise to runaway solutions. In our cosmological context, rather than a problem this can actually be a virtue,  because a phase of accelerated expansion is in a sense  an instability of the classical evolution. Indeed, ghosts have been suggested as models of phantom dark energy~\cite{Caldwell:1999ew,Carroll:2003st}. The real trouble is that, at the quantum level, a ghost corresponds to a particle with negative energy and induces a decay of the vacuum, through processes in which the vacuum decays into ghosts plus normal particles, and  renders the theory inconsistent. 

The main purpose of this paper is to discuss and clarify some subtle conceptual issues related to  this apparent ghost-like degree of freedom
and to show that, in fact, in this theory there is no propagating ghost-like degree of freedom.
The paper is organized as follows. In sect.~\ref{sect:eff} we show that the status of a non-local equation such as \eq{modela2} is that of an effective classical equation, derived from some classical or quantum averaging in a more fundamental theory.
In sect.~\ref{sect:vacsta} we show that similar apparent ghosts even appear in massless GR when one decomposes the metric perturbation into a transverse-traceless 
part $\hmn^{\rm TT}$ and a trace part $\emn s$. This is due to the fact that the relation between $\{\hmn^{\rm TT},s\}$ and the original metric perturbations $\hmn$ is non-local. We show that (contrary to some statements in the literature) the apparent ghost field $s$ is not neutralized by the helicity-0 component of $\hmn^{\rm TT}$. Rather, what saves the vacuum stability of GR, in these variables, is that $s$ (as well as the helicity-0, $\pm 1$ components of $\hmn^{\rm TT}$), is a non-propagating field and cannot be put on the external lines, nor in loops. Beside having an intrinsic conceptual interest, this analysis will also show that the same considerations extend straightforwardly to the non-local modification of GR that we are studying.
Finally, in sect.~\ref{sect:aux} we will work out the explicit relation between the fake ghost that is suggested by \eq{TT}, and the spurious degrees of freedom that are know to emerge when a non-local theory is written in local form by introducing auxiliary fields. Sect.~\ref{sect:concl} contains our conclusions.

\section{Non local QFT or  classical effective equations?}\label{sect:eff}

A crucial point of \eq{modela2}, or of its linearization (\ref{line1}), is that they contain explicitly a retarded propagator. This retarded prescription is forced by causality, which we do not want to give up. We are used, of course, to the appearance of retarded propagators in the solutions of classical equations. Here however the retarded propagator already appears  in the equation itself, and not only in its solution. Is it possible to obtain such an equation from a variational principle? The answer, quite crucially, is no.
As already observed by various authors
\cite{Deser:2007jk,Barvinsky:2011rk,Jaccard:2013gla},  a retarded inverse d'Alembertian cannot be obtained from the variation of a non-local action.
Consider for illustration a non-local term in an action of the form
$\int dx \phi\iBox\phi$, where $\phi$ is some scalar field, and $\iBox$ is defined with respect to some Green's function $G(x;x')$. Taking the variation with respect to $\phi(x)$ we get
\bees
&&\frac{\d}{\d\phi(x)}\int dx' \phi(x') (\iBox\phi )(x')=
\frac{\d}{\d\phi(x)} \int dx' dx'' \phi(x') G(x';x'') \phi(x'')\nn\\
&&=\int dx' [G(x;x')+G(x';x)] \phi(x')\, . \label{symGreen}
\ees
We see that the variational of the action automatically symmetrizes the Green's
function. It is therefore impossible to obtain in this way a retarded Green's function in the equations of motion, since $G_{\rm ret}(x;x')$ is not symmetric under $x\leftrightarrow x'$;  rather 
$G_{\rm ret}(x';x)=G_{\rm adv}(x;x')$. 
The same happens if we take the variation of the Lagrangian (\ref{Lquadr}).  
Writing explicitly the convolution with the Green's function as we did in \eq{symGreen} we find that it is not possible to get the term
$P_{\rm ret}^{\mu\nu}P_{\rm ret}^{\rho\sigma}$ in \eq{line1}. If in the action the term $\iBox$ that appears in $P^{\mu\nu}$ is defined with a symmetric Green's function, so that $G(x;x')=G(x';x)$, we find the same Green's function in
the equation of motion. If, in contrast, we use 
$\hmn P_{\rm ret}^{\mu\nu}P_{\rm ret}^{\rho\sigma}\hrs$ in the action, in the equation of motions we get $(P_{\rm ret}^{\mu\nu}P_{\rm ret}^{\rho\sigma}+
P_{\rm adv}^{\mu\nu}P_{\rm adv}^{\rho\sigma})\hrs$.

Of course, one can take the point of view that the classical theory is defined by its equations of motion, while the
action is simply a convenient ``device" that, through a set of well defined rules, allows us to compactly summarize   the equations of motion. We can then take the formal variation of the action and at the end replace by hand all factors $\iBox$ by
$\iBox_{\rm ret}$ in the equation of motion. 
This is  indeed the 
procedure used in \cite{Soussa:2003vv,Deser:2007jk}, in the context of non-local  gravity theories with a Lagrangian of the form $Rf(\iBox R)$. As long as we see the Lagrangian as a ``device" that, through a well defined procedure, gives a classical equation of motion, this prescription is certainly legitimate. However, any connection between these classical causal equations of motion and  the {\em quantum} field theory described by such a Lagrangian is now lost. 
In particular, 
the terms in \eq{prop} or in \eq{TT} that apparently describe the exchange of a healthy massless scalar plus a ghostlike massive scalar are just the propagators that, to reproduce  \eq{line1}, 
after the variation must be set equal to retarded propagators. Taking them  as Feynman propagators in a QFT gives a quantum theory
that  has nothing to do with our initial classical equation
(\ref{line1}) and that has  dynamical degrees of freedom that, with respect to our original problem, are  spurious.  

Thus, \eq{modela2} is not the classical equation of motion of a non-local quantum field theory. To understand its conceptual meaning, we observe that 
non-local equations involving the retarded propagator appear in various situation in physics, but are never fundamental. They rather  typically emerge after performing some form of averaging, either purely classical or at the quantum level. 
In particular, non-local field equations govern the effective dynamics of the vacuum expectation values of quantum fields, which include the quantum corrections to the effective action. The standard path integral approach provides the dynamics for  the in-out matrix element of a quantum field, e.g. 
$\langle 0_{\rm out}|\hat{\phi}|0_{\rm in}\rangle$ or, in a semiclassical approach to gravity,  
$\langle 0_{\rm out}|\hat{g}_{\mu\nu}|0_{\rm in}\rangle$. The classical equations for these quantities are however determined by the Feynman propagator, so they are not causal, since they contain both the retarded and the advanced Green's function.  This is not surprising, since the in-out matrix element are not directly measurable quantities, but only provide intermediate steps in the QFT computations.
Furthermore, even if $\hat{\phi}$ is a hermitean operator, its in-out matrix element are complex. In particular, this makes it impossible to interpret 
$\langle 0_{\rm out}|\hat{g}_{\mu\nu}|0_{\rm in}\rangle$ as an effective metric. 
In contrast, the in-in matrix elements are real, and satisfy non-local but causal 
equations  \cite{Jordan:1986ug,Calzetta:1986ey}, involving only  retarded propagators (that can be computed using the Schwinger-Keldysh formalism).

Similar non-local but causal equations can  also emerge from a purely classical averaging procedure, when one separates the dynamics of a system into a long-wavelength and a short-wavelength part. One can then obtain an effective non-local but causal equation for the long-wavelength modes by integrating out the short-wavelength modes, see e.g. \cite{Carroll:2013oxa} for a recent example in the context of cosmological perturbation theory. 
Another purely classical example comes from the standard post-Newtonian/post-Minkowskian formalisms for GW production  \cite{Blanchet:2006zz,Maggiore:1900zz}.  In linearized theory the gravitational wave (GW) amplitude $\hmn$ is determined by  
$\Box\bhmn=-16\pi G\Tmn$, where $\bhmn=\hmn-(1/2)h\emn$.
In such a radiation problem this equation is solved with the retarded Green's function,  
$\bhmn=-16\pi G\iBox_{\rm ret}\Tmn$. When the non-linearities of GR are included,
the GWs generated at some perturbative order become themselves sources for the GW generation at the next order. In the far-wave zone, this iteration gives rise to effective equations for $\bhmn$ involving $\iBox_{\rm ret}$.

In summary, non-local equations involving $\iBox_{\rm ret}$ are {\em not} the classical equation of motion a non-local QFT (a point already made e.g. in
\cite{Tsamis:1997rk,Deser:2007jk,Barvinsky:2011rk,Deser:2013uya,Ferreira:2013tqn}). Even if we can find a classical Lagrangian whose variation reproduces them (once supplemented with the $\iBox\ra \iBox_{\rm ret}$ prescription after having performed the variation), the {\em quantum} field theory described by this Lagrangian has a priori nothing to do with the problem at hand. Issues of quantum consistencies (such as the possibility of a vacuum decay amplitude induced by ghosts) can only be addressed in the fundamental theory that, upon classical or quantum smoothing, produces these non-local (but causal) classical equations.

So, there is no sense, and no domain of validity, in which the Lagrangian
(\ref{Lquadr})  can be used to define a QFT associated to  our theory.
To investigate whether the classical equation (\ref{modela2}) derives from a QFT with a stable quantum vacuum we should identify the fundamental theory and the smoothing procedure that give rise to it,  and only in this framework we can pose the 
question.

\section{Vacuum stability in massless and massive gravity}\label{sect:vacsta}

\subsection{A fake ghost in massless GR}

It is instructive to see more generally how spurious degrees of freedom, and in particular spurious ghosts, can appear when one uses  non-local variables. A simple and quite revealing example is provided by GR itself. We have already discussed this example in app.~B of \cite{Jaccard:2012ut}, but it is useful to re-examine and expand it in this context. We consider GR linearized over flat space. The quadratic Einstein-Hilbert action is
\be\label{Squadr}
S_{\rm EH}^{(2)}=\frac{1}{2}\int d^{d+1}x \,
\hmn{\cal E}^{\mu\nu,\rho\sigma}\hrs
\ee
We decompose  the metric as
\be\label{decomphmn}
\hmn =\hmn^{\rm TT}+\frac{1}{2}(\pam \eps_{\nu}+\pan \eps_{\mu}) +\frac{1}{d}\emn s\, ,
\ee
where $\hmn^{\rm TT}$ is transverse and traceless, $\paM \hmn^{\rm TT}=0$, 
$\eMN \hmn^{\rm TT}=0$. The vector $\eps_{\mu}$ could be further decomposed as 
$\eps_{\mu}=\eps_{\mu}^{\rm T}+\pam\alpha$, where
$\paM \eps_{\mu}^{\rm T}=0$. Under a linearized diffeomorphism
$\hmn\ra\hmn -(\pam\xin+\pan\xim)$ we have $\eps_{\mu}\ra\eps_{\mu}-\xi_{\mu}$ while the tensor $\hmn^{\rm TT}$ and the scalar $s$ are gauge invariant.
Plugging \eq{decomphmn} into \eq{Squadr} we find that $\eps_{\mu}$ cancels (as it is obvious from the fact that \eq{Squadr} is invariant under linearized diffeomorphisms and $\eps_{\mu}$ is a pure gauge mode), and
\be\label{SEH2nl}
S_{\rm EH}^{(2)}=\frac{1}{2}\int d^{d+1}x \,\[\hmn^{\rm TT}\Box (h^{\mu\nu })^{\rm TT}
-\frac{d-1}{d}\, s\Box s\]\, .
\ee
Performing the same decomposition in the energy-momentum tensor, the interaction term 
can be written as
\be\label{Sint2}
S_{\rm int}=\frac{\kappa}{2}\int d^{d+1}x\, \hmn\TMN 
=\frac{\kappa}{2}\int d^{d+1}x\,\[ \hmn^{\rm TT}(\TMN)^{\rm TT}+\frac{1}{d}sT\]
\, ,
\ee
so the equations of motion derived from $S_{\rm EH}^{(2)}+S_{\rm int}$ are
\bees
\Box\hmn^{\rm TT}&=&-\frac{\kappa}{2}\Tmn^{\rm TT}\, ,\label{BoxhTT}\\
\Box s&=&\frac{\kappa}{2(d-1)}T\, .\label{Boxs}
\ees
This result can be surprising, because it seems to suggest that in ordinary massless GR we have many more propagating degrees of freedom than expected: the components of the transverse-traceless tensor $\hmn^{\rm TT}$ (i.e. 5 components in $d=3$) plus the scalar $s$. Note that  these degrees of freedom are gauge invariant, so they cannot be  gauged away. Furthermore, from \eq{SEH2nl}
the scalar $s$ seems a ghost! 
Of course these conclusions are wrong, and for any $d$ linearized GR is a ghost-free theory; in particular, in $d=3$ it only  has two radiative degrees of freedom, corresponding to the $\pm 2$ helicities of the graviton (and, in generic $d$,  $(d+1)(d-2)/2$ radiative degrees of freedom, corresponding to the fact that the little group is $SO(d-1)$). We know very well that the remaining degrees of freedom of GR are physical (i.e.  gauge-invariant) but non-radiative. How is this consistent with the fact that $s$, as well as all the extra components of $\hmn^{\rm TT}$, satisfy a Klein-Gordon rather than a Poisson equation?

The answer, as discussed in \cite{Jaccard:2012ut}, is related to the fact that $\hmn^{\rm TT}$ and $s$ are non-local functions of the original metric perturbation $\hmn$. In particular, inverting \eq{decomphmn} one finds that
\be\label{defshmn}
s=\(\eMN-\frac{1}{\Box}\paM\paN\)\hmn=P^{\mu\nu}\hmn\, .
\ee 
The fact that $s$, as a function of $\hmn$, is non-local in time means that the initial data assigned on $\hmn$ on a given time slice are not sufficient to evolve $s$, so a naive counting of degrees of freedom goes wrong.  A simple but instructive example of what exactly goes wrong is provided by a scalar field $\phi$ that satisfies a Poisson equation  $\n^2\phi=\rho$. If we  define a new field $\tilde{\phi}$ from  $\tilde{\phi}=\iBox\phi$,  the original  Poisson  equation   can be rewritten as  
\be\label{eqtildephi}
\Box\tilde{\phi}=\n^{-2}\rho\equiv\tilde{\rho}\, , 
\ee
so now $\tilde{\phi}$ looks like a propagating degree of freedom. However, for $\rho=0$ our original equation $\n^2\phi=\rho$ only has the solution $\phi=0$. If we want to rewrite it in terms of $\tilde{\phi}$ without introducing spurious degrees of freedom we must therefore supplement \eq{eqtildephi} with the condition that, when $\rho=0$, $\tilde{\phi}=0$. In other words,  the homogeneous plane wave solutions of \eq{eqtildephi},
\be\label{tildephiaa}
\tilde{\phi}_{\rm hom}(x)=\int \frac{d^3\vk}{(2\pi)^3}\,
\[ a_{\vk} e^{ikx} +a_{\vk}^{*} e^{-ikx}\]
\ee
is fixed uniquely by the original equation, and $a_{\vk}, a_{\vk}^{*}$ cannot be considered as free parameters that,
upon quantization, give rise to the creation and annihilation operators of the quantum theory. 

Exactly the same situation takes place in GR, for the field $s$ and for the the extra components in $\hmn^{\rm TT}$. For instance, writing  $s$ 
in terms of the variables entering the $(3+1)$ decomposition (and specializing   to $d=3$) one finds
\be\label{sPhiPsi}
s=6\Phi- 2\iBox\n^2(\Phi+\Psi)\, ,
\ee
where $\Phi$ and $\Psi$ are the scalar Bardeen's variable defined in flat space (see \cite{Jaccard:2012ut}). Since $\Phi$ and $\Psi$ are non-radiative and satisfy Poisson equations,  $s$ is non-radiative too, and it is just the $\iBox$ factor in \eq{sPhiPsi} that (much as the $\iBox$ in the definition of $\tilde{\phi}$), potentially introduces a fake propagating degree of freedom. In order to eliminate such  a spurious degree of freedom, we must supplement
\eq{Boxs} with the condition that $s=0$ when $T=0$, i.e. we must discard again the homogeneous solution of \eq{Boxs} (and similarly for \eq{BoxhTT}).
This implies that, at the quantum level, there are no creation and annihilation operators associated to $s$. Therefore $s$ cannot appear on the external legs of a Feynman diagram, and there is no Feynman propagator associated to it, so it cannot circulate in the loops.

\begin{figure}[t]
\begin{center}
\includegraphics[width=0.2\textwidth]{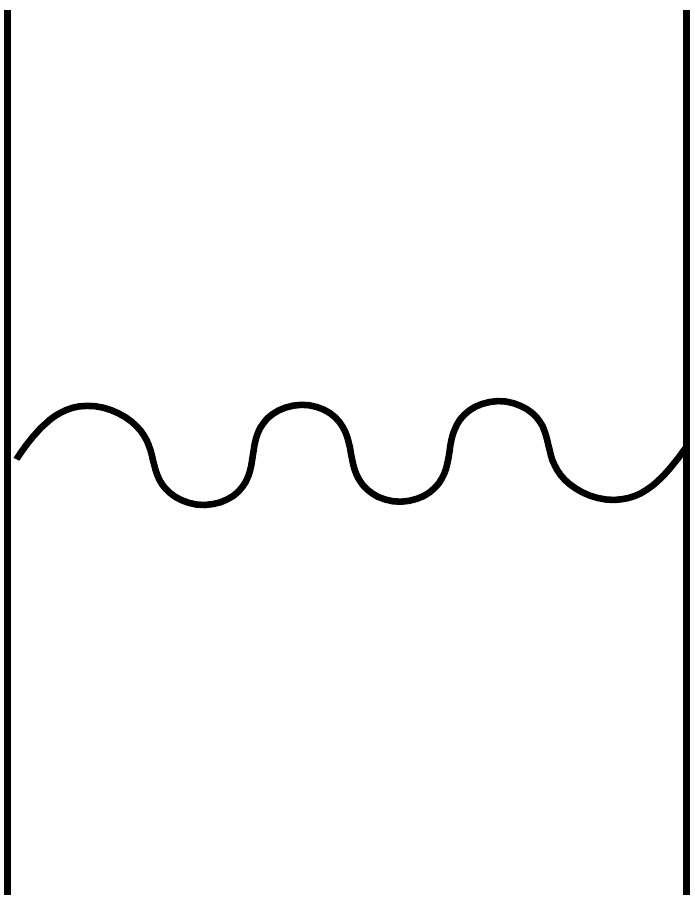}
\caption{\label{fig:tree} The tree-level Feynman graph describing 
the exchange of a graviton between two matter lines. In this graph, the contribution due to the exchange of the scalar $s$ is canceled by the exchange of the helicity-0 component of
$\hmn^{\rm TT}$.}
\end{center}
\end{figure}
 
\begin{figure}[t]
\begin{center}
\includegraphics[width=0.6\textwidth]{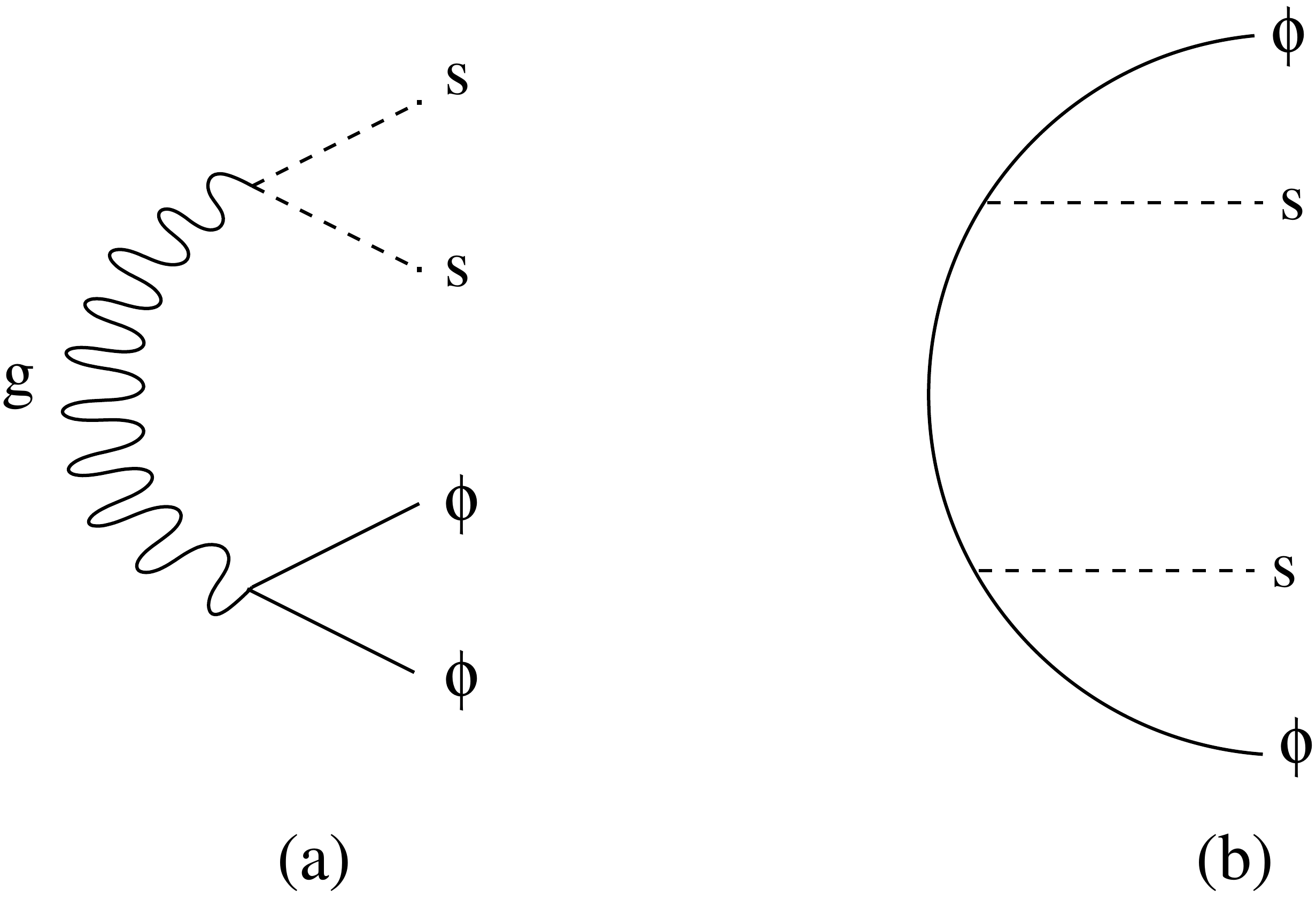}\hspace{3cm}
\caption{\label{fig:vacdecay}Examples of vacuum decay graphs that would be induced by $s$ if we were allowed to put it on the external legs. The wavy lines   denotes gravitons,  the solid line a generic  matter field $\phi$ (or a graviton itself) and the dashed line the would-be ghost field $s$.}
\end{center}
\end{figure}

One might also observe that the contribution to the propagator of $s$ is canceled by an equal and opposite contribution due to the helicity-0 component of 
$\hmn^{\rm TT}$, 
see app.~A1 of \cite{Hassan:2011tf}. However this only shows that, in the classical matter-matter interaction described by tree level diagrams such as that in Fig.~\ref{fig:tree},
these contributions cancel and we remain, as expected, with the contribution from the exchange of the helicity $\pm 2$. This cancellation has nothing to do with the vacuum stability  of GR. Consider in fact the graphs shown in Fig.~\ref{fig:vacdecay}. If $s$ were a dynamical ghost field that can be put on the external line, these graphs would describe a vacuum decay process. Such a process is kinematically allowed because the ghost $s$ carries a negative energy that compensates the positive energies of the other final particles. There is no corresponding vacuum decay graph in which we replace $s$ by the helicity-0 component of
$\hmn^{\rm TT}$, since the latter is not a ghost, and the process is no longer kinematically allowed. In any case, these processes have different final states, so the positive probability for, e.g., the decay ${\rm vac}\ra ss\phi\phi$ shown in 
Fig.~\ref{fig:vacdecay} (where $\phi$ is any normal matter field, or a graviton) cannot be canceled by anything.

\begin{figure}[t]
\begin{center}
\includegraphics[width=0.8\textwidth]{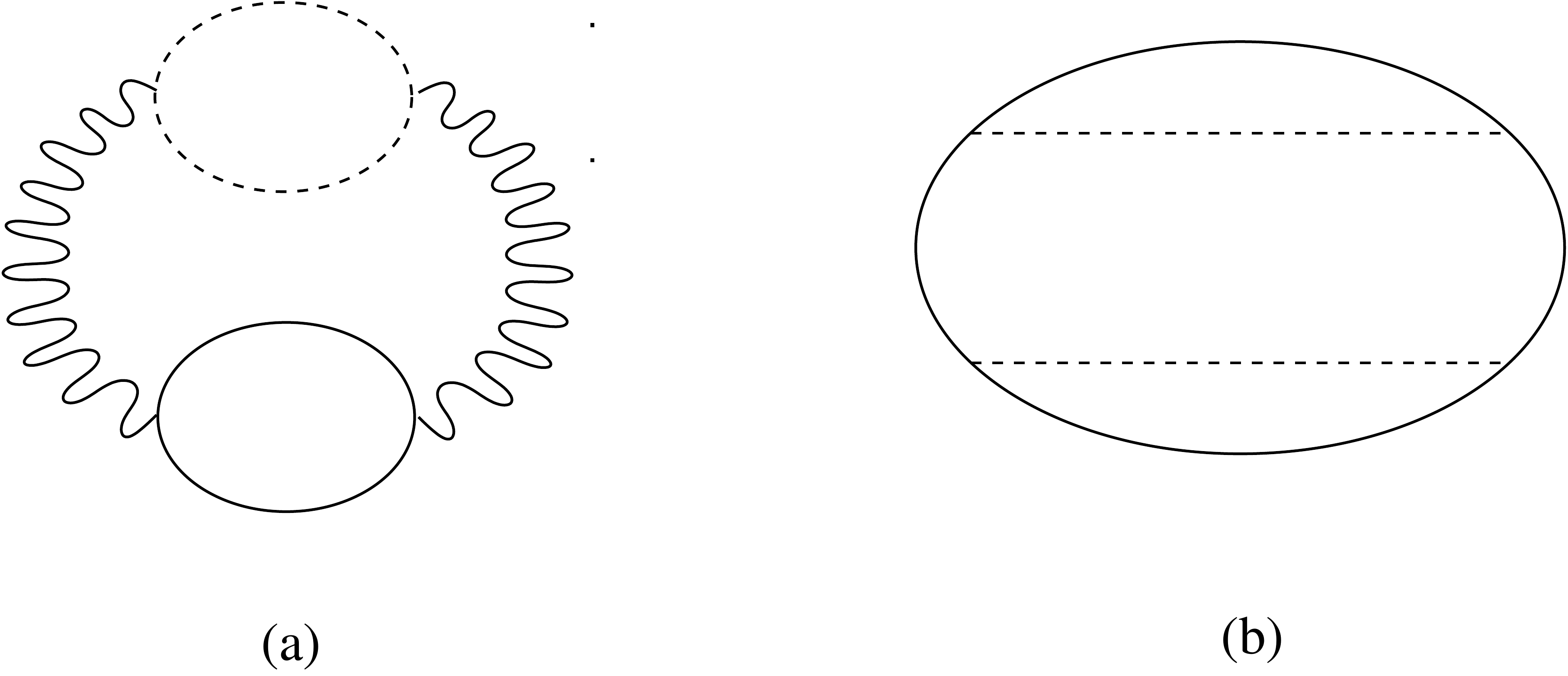}\hspace{3cm}
\caption{\label{fig:vacvac}The vacuum-to-vacuum  diagrams corresponding to the processes shown in Fig.~\ref{fig:vacdecay}.}
\end{center}
\end{figure}

It is interesting, and somewhat subtle, to understand the same point in terms of the imaginary part of vacuum-to-vacuum diagrams. The vacuum-to-vacuum diagrams corresponding to the processes of  Fig.~\ref{fig:vacdecay} are shown in
Fig.~\ref{fig:vacvac}. Here, whenever we have a dashed line corresponding to $s$, we indeed have a corresponding graph where this line is replaced by the propagation of the 
helicity-0 component of  $\hmn^{\rm TT}$, and one might believe that these graphs cancel. In fact this is not true, due to a subtlety in the $i\eps$ prescriptions of the propagators. For a normal particle  the usual scalar propagator is $-i/(k^2+m^2-i\eps)$ (with our $(-,+,+,+)$ signature). For a propagating ghost the correct prescription is instead $i/(k^2-m^2+ i\eps)$. 
As discussed in \cite{Cline:2003gs}, this  $+i\eps$  choice propagates negative energies forward in time but preserves the unitarity of the theory and the optical theorem. With a  $-i\eps$  choice, in contrast, ghosts carry positive energy but negative norm, and the probabilistic interpretation of QFT is lost. This latter choice is therefore unacceptable. 
In our case $m=0$ and the sum of the contributions to each internal line of the ``healthy" helicity-0 component of  $\hmn^{\rm TT}$ and the ghost $s$ is
\be\label{ieps}
-\frac{i}{k^2-i\eps}+\frac{i}{k^2+i\eps}
\ee
We see that, because of the different $i\eps$ prescriptions, these two terms do not cancel. Indeed, the ghost contribution to the diagrams of Fig.~\ref{fig:vacvac} generates an imaginary part that corresponds to the modulus square of the corresponding diagrams in Fig.~\ref{fig:vacdecay}, as required by unitarity. In contrast,  the contribution to the diagram of Fig.~\ref{fig:vacvac} from the helicity-0 component of  $\hmn^{\rm TT}$ has no imaginary part, again in agreement with unitarity, since the processes corresponding to Fig.~\ref{fig:vacdecay}, with $s$ replaced by the helicity-0 component of  $\hmn^{\rm TT}$, are not kinematically allowed.

To sum up, what saves the vacuum stability in GR  is not a cancelation between  the contributions of the ghost $s$ and that  of the helicity-0 component of
$\hmn^{\rm TT}$. If one treats them as  propagating degrees of freedom there is no such cancelation, and one reaches the (wrong) conclusion that in GR the vacuum is unstable.
Rather, vacuum stability is preserved by the fact that the field $s$, as well as the extra components of  $\hmn^{\rm TT}$,
are non-radiative. There are no destruction nor creation operators associated to them,   and we are not allowed  to put these fields on the external lines or in  loops.

In other words, the theory defined by \eq{SEH2nl} is not  equivalent to 
that defined by the quadratic Einstein-Hilbert action  (\ref{Squadr}), because the non-local transformation between  $\hmn$ and  $\{\hmn^{\rm TT},s\}$ introduces spurious  propagating modes. We can still describe GR using the formulation
in terms of $\{\hmn^{\rm TT},s\}$, but in this case we must impose 
on \eq{Boxs}  the boundary condition that $s=0$ when $T=0$ (and similarly for the extra components of $\hmn^{\rm TT}$
 in \eq{BoxhTT}), in order to eliminate these spurious modes.

\subsection{The apparent ghost in the  non-local massive theory}
 
It is now straightforward to make contact  between linearized GR and its non-local 
massive deformation given by \eq{line1}. Integrating by parts the operator $P^{\mu\nu}$ and using \eqs{SEH2nl}{defshmn}, the Lagrangian (\ref{Lquadr}) can be written as 
\bees
{\cal L}_2&=&\frac{1}{2}\hmn{\cal E}^{\mu\nu,\rho\sigma}\hrs
-\frac{d-1}{2d}\, m^2 (P^{\mu\nu}\hmn)^2\nn\\
&=&\frac{1}{2}\,\[\hmn^{\rm TT}\Box (h^{\mu\nu })^{\rm TT}
-\frac{d-1}{d}\, s(\Box +m^2)s\]\, .\label{Lquadr2}
\ees
Thus, the non-local term in \eq{Lquadr} is simply  a mass term for the field $s$. However, in the original equation of motion (\ref{line1}) that this action is supposed to reproduce, the non-local term was defined with  the retarded Green's function. Thus, in order not to introduce spurious propagating degrees of freedom, we must simply continue to impose the condition that $s$ is a non-radiative field, just as we did in GR. In other words, now the equation of motion (\ref{Boxs}) is replaced by 
\be\label{Boxs2}
(\Box +m^2)s=\frac{\kappa}{2(d-1)}T\, ,
\ee
and, just as in \eq{tildephiaa}, we must refrain from interpreting the coefficients 
of the the plane waves $e^{\pm ikx}$ with $k^2=m^2$ as free parameters that, upon quantization, give rise to creation and annihilation operators. Thus, again, there are no creation and annihilation operators associated to $s$, which therefore cannot appear on the external lines of graphs such as those in Fig.~\ref{fig:vacdecay}, not in the internal lines of
graphs such as those in Fig.~\ref{fig:vacvac}, and there is no vacuum decay. Observe that, since the pole of $s$ is now massive while that of the helicity-0 mode 
of $\hmn^{\rm TT}$
remain massless, there is no cancelation among them in the tree graphs that describe the {\em classical} matter-matter interaction, which is therefore modified at cosmological distances, compared to GR. This is just as we want, since our aim is to modify classical gravity in the IR. In contrast, the lack of cancelation between $s$ and the helicity-0 mode has nothing to do with unitarity and vacuum decay. As discussed above, this cancellation does not take place even in the $m=0$ case. Graphs such as those in
Fig.~\ref{fig:vacdecay} could not be canceled by anything, and the reason why the vacuum decay amplitude in GR is zero is that these graphs simply do not exist, because we cannot put $s$ on the external lines.\footnote{Observe that, because of the lack of cancelation due to the signs of the $i\eps$ factors in \eq{ieps}, the argument proposed in sect.~6.1 of \cite{Jaccard:2013gla} (where the contribution to the vacuum-to-vacuum amplitude coming from  the massive ghost and  the massless scalars in \eq{TT} 
partially canceled, modulo corrections ${\cal O}(m^2/E^2)$), is incorrect. However, we now see that the result is even stronger. The vacuum decay amplitude in the theory with finite $m$ is not suppressed by factors $m^2/\Lambda^2$ (where $\Lambda$ is the UV cutoff) but is in fact identically zero, because $s$ cannot appear on the external lines of a Feynman diagram, neither for $m=0$ nor for $m\neq 0$.}

It is instructive to compare the situation with the usual local theory  of linearized massive gravity. With a generic local mass term, the quadratic Lagrangian reads
\be\label{calLFP1}
{\cal L}_{\rm FP}= \frac{1}{2}
\hmn {\cal E}^{\mu\nu,\rho\sigma}\hrs
- \frac{m^2}{2} (b_1\hmn\hMN +b_2h^2  )\, .
\ee
Using again the decomposition (\ref{decomphmn}), now the action depends also on $\eps^{\mu}$, since the invariance under linearized diffeomorphisms is broken. Writing 
$\eps_{\mu}=\eps_{\mu}^{\rm T}+\pam\alpha$, the scalar sector now depends both on $s$ and $\a$, with $\hmn^{\rm scalar}=\pam\pan\a +(1/d)\emn s$.
In particular, the mass term in \eq{calLFP1} produces a term proportional to $(b_1+b_2)(\Box\alpha)^2$. For $b_1,b_2$ generic, this higher-derivative term  gives rise to a ghost, and the Fierz-Pauli tuning $b_1+b_2=0$ is designed so to get rid of it. Indeed, this longitudinal mode of the metric of the form $\pam\pan\a$
is nothing but the mode that is isolated using  the \Stu formalism, and the dRGT theory \cite{deRham:2010ik,deRham:2010kj}
 is constructed just to
ensure that its equations of motion remains of second order, even at the non-linear level.
The situation is quite different from that in \eq{Lquadr2}, where no higher-derivative term is generated, but we just added a mass term to an already non-radiative field.

\section{Spurious degrees of freedom from auxiliary fields}\label{sect:aux}

An alternative way of studying the degrees of freedom   of a non-local theory 
is to transform it into a local theory by introducing auxiliary fields, see e.g.
\cite{Nojiri:2007uq,Jhingan:2008ym}.
As it has been recognized in various recent papers \cite{Koshelev:2008ie,Koivisto:2009jn,Barvinsky:2011rk,Deser:2013uya}, such  ``localization" procedure introduces however spurious solutions, and in particular spurious ghosts. This is in fact an equivalent way of understanding that the apparent ghosts of these theories do not necessarily correspond to propagating degrees of freedom. An example that has been much studied is the non-local model originally proposed by Deser and Woodard~\cite{Deser:2007jk}, which is based on the action
\be\label{SDW}
S=\frac{1}{16\pi G}\int d^4x \sqrt{-g} R \[ 1+f(\iBox R)\]\, ,
\ee
for some function $f$. This can be formally rewritten in local form introducing two fields $\xi(x)$ and $\phi(x)$ and writing
\be
S=\frac{1}{16\pi G}\int d^4x \sqrt{-g} \left\{ 
R \[ 1+f(\phi)\] +\xi (\Box\phi -R)\right\}\, .
\ee
Thus, $\xi$ is a Lagrange multiplier that enforces the equation $\Box\phi=R$, so that formally $\phi=\iBox R$. The kinetic term $\xi\Box\phi=-\pam\xi\paM\phi$ can be diagonalized writing
$\xi=\varphi_1+\varphi_2$, $\phi=\varphi_1-\varphi_2$, and then 
\be\label{SDWloc}
S=\frac{1}{16\pi G}\int d^4x \sqrt{-g} \left\{ 
R \[ 1+f(\varphi_1-\varphi_2) -\varphi_1-\varphi_2\]-\pam\varphi_1\paM\varphi_1
+\pam\varphi_2\paM\varphi_2
\right\}\, ,
\ee
and we see that one of the two auxiliary fields ($\varphi_2$, given our signature) is a ghost. However, this apparent ghost is a spurious degree of freedom, as it is immediately understood observing 
that the above formal manipulation hold even when 
 the function $f(x)$ is equal to a constant $f_0$ \cite{Koshelev:2008ie} (or, in  fact,  even when $f=0$). In this case the original action (\ref{SDW}) is obviously the same as GR with a rescaled Newton constant, and certainly has no ghost (and, in fact, it has no ghost also for a broad class of functions $f(\iBox R)$ \cite{Deser:2013uya}).
Once again, the point is that \eq{SDWloc} is equivalent to \eq{SDW} only if
we  discard the homogeneous solution of $\Box\phi=R$, and therefore there are no annihilation and creation operators associated to $\phi$ (nor to $\xi$).
A similar example has been given,  for a non-local model 
based on the term $R^{\mu\nu}\iBox\Gmn$, in  \cite{Barvinsky:2011rk}, where it was also clearly recognized that the auxiliary ghost field that results from the localization procedure never exists as a propagating degree of freedom, and does not appear in the external lines of the Feynman graphs.

Exactly the same happens in our model. To {\em define} the model we must specify what $\iBox$ actually means. In general, an equation such as $\Box U=-R$ is solved by
\be
U=-\iBox R =U_{\rm hom}(x)-\int d^{d+1}x'\, \sqrt{-g(x')}\, G(x;x') R(x')\, ,
\ee
where $U_{\rm hom}(x)$ is any solution of $\Box U_{\rm hom}=0$ and $G(x;x')$ is any a Green's function of the $\Box$ operator. To define our model we must specify what definition of $\iBox$ we use, i.e. we must specify the Green's function and the solution of the homogeneous equation. In our case we use the retarded Green's function, but still we must complete the definition of $\iBox$ by specifying $U_{\rm hom}(x)$. A possible choice is $U_{\rm hom}(x)=0$. Then, in \eq{modela2},
\be\label{defiB}
(\iBox R)(x) \equiv \int d^{d+1}x'\, \sqrt{-g(x')}\,  G_{\rm ret}(x;x') R(x')\, .
\ee
A similar choice must be made in the non-local operators which enter in the extraction of the transverse part in \eq{modela2}. Thus, at the linearized level, with this definition of $\iBox$, in \eq{line1} we have
\be
P_{\rm ret}^{\rho\sigma}\hrs \equiv h(x)-
\int d^{d+1}x'\,  G_{\rm ret}(x;x') (\paR\paS\hrs)(x')\, ,
\ee
and similarly
\bees
P_{\rm ret}^{\mu\nu}P_{\rm ret}^{\rho\sigma}\hrs
&\equiv&\eMN \[  h(x)-
\int d^{d+1}x'\,  G_{\rm ret}(x;x') (\paR\paS\hrs)(x')\]\\
&&-
\paM\paN \int dx' G_{\rm ret}(x;x') \[ h(x')-
\int d^{d+1}x'\,  G_{\rm ret}(x';x'') (\paR\paS\hrs)(x'')\]\, .\nn
\ees
Consider now what happens if we rewrite the theory in a local form, introducing
$U=-\iBox R$ and 
$\Smn=-U\gmn$.
Formally, \eq{modela2} can be written as 
\be\label{loc1}
\Gmn -\frac{d-1}{2d}\, m^2\Smn^{\rm T}=8\pi G\,\Tmn\, ,
\ee
where $S_{\mu\nu}=
S_{\mu\nu}^{\rm T}+\frac{1}{2}(\n_{\mu}S_{\nu}+\n_{\nu}S_{\mu})$. To make contact with \eq{line1} we linearize over flat space and we use \eq{Tflat}. Then
\eq{loc1} can be rewritten as the coupled system
\bees\label{line1U}
{\cal E}^{\mu\nu,\rho\sigma}\hrs
-\frac{d-1}{d}\, m^2 P_{\rm ret}^{\mu\nu}U
&=&-16\pi G\TMN\, ,\label{EPU}
\\
\Box U&=&-R\, .\label{RBoxU}
\ees
Such a local form of the equations can be convenient, particularly for numerical studies, because it transforms the original integro-differential equations into a set of coupled differential equations. However, exactly as in the example discussed above, 
it introduces spurious solutions.  The choice of homogeneous solution, that in
the original non-local formulation amounts to a definition of the theory, is now translated into a choice of initial conditions on the field $U(x)$. There is one, and only one choice, that gives back our original models. For instance, if the original non-local theory is defined through \eq{defiB}, we must choose the initial conditions on $U$ in
\eq{RBoxU} such that the solution of the associated homogeneous equation vanishes. 
In any case, whatever the choices made in the definition of $\iBox$, the corresponding homogeneous solution of \eq{RBoxU} is fixed, and does not represent a free field that we can take as an extra degree of freedom of the theory. In flat space this homogeneous solution is a superposition of plane waves of the form (\ref{tildephiaa}), and the coefficients $a_{\vk},a_{\vk}^*$ are fixed by the definition of $\iBox$ (e.g. at the value $a_{\vk}=a_{\vk}^*=0$ if we use the definition (\ref{defiB})~), and at the quantum level it makes no sense to promote them to annihilation and creation operators. There is no quantum degree of freedom associated to them.

Comparing \eq{EPU} with \eq{line1} we see that, at the linearized level, that $U=P_{\rm ret}^{\rho\sigma}\hrs$. Therefore at the linearized level $U$ is the same as the variable $s$ given in \eq{defshmn}. The fact that the  homogeneous solutions for $U$ does not represent a free degree of freedom means that the same holds for  $s$. We therefore reach the same conclusion of the previous section: the homogeneous solutions for $s$ do not describe propagating degrees of freedom, and at the quantum level there are no creation and annihilation operators associated to it. 

\section{Conclusions}\label{sect:concl}

Non-local modifications of GR have potentially very interesting cosmological consequences. At the conceptual level, they raise however some issues of principle which must be understood before using them confidently to compare with cosmological observations. 
In particular, these equations feature the {\em retarded} inverse of the d'Alembertian. The retarded prescription ensures causality, but at the same time the fact $\iBox_{\rm ret}$ appears not only in the solution of such equations, but already in the equations themselves,  tells us that such equations cannot be fundamental. Rather, they are effective classical equations.

Such non-local effective equations can emerge in a purely classical context. Typical examples are obtained when   integrating out the short-wavelength  modes to obtain an effective theory for 
long-wavelength  modes. Another example is given by the formalism for gravitational-wave production, beyond leading order. In both cases one basically re-injects a retarded solution, obtained to lowest order, into the equation governing the next-order corrections. Another way to obtain non-local equations  is by performing a quantum averaging, in particular when working with  the in-in expectation values of the quantum fields, and deriving these equations from an effective action that  takes into account the radiative corrections. In particular, in semiclassical quantum gravity we can write such effective non-local (but causal) equations for an effective metric $\langle 0_{\rm in}|\hat{g}_{\mu\nu}|0_{\rm in}\rangle$.

We have seen (in agreement with various recent works, e.g. \cite{Koshelev:2008ie,Koivisto:2009jn,Barvinsky:2011rk,Deser:2013uya}) that, if one is not careful, it is quite easy to introduce spurious degrees of freedom in these models, which furthermore are ghost-like. Basically, this originates from 
the fact that the kernel of the $\iBox$ operator is non-trivial: the equation $\iBox(0)=f$ does not imply that $f=0$ but only that $f$ satisfies $\Box f=0$. 
The non-local equations that we are considering only involve the retarded solutions of equations of the form $\Box f=j$, for some source $j$, i.e.
\be
f(x)=\int dx' G_{\rm ret}(x;x') j(x')\, .
\ee
However, any action principle that (with some more or less formal manipulation, as discussed in sect.~\ref{sect:eff}) reproduces the equation $\Box f=j$ will automatically carry along the most general solution of this equation, of the form
\be
f(x)=f_{\rm hom}+\int dx' G(x;x') j(x')\, ,
\ee
where $\Box f_{\rm hom}=0$ and $G(x;x')$ is a generic Green's function. In order to recover the solutions that actually pertain to our initial non-local theory we must 
impose the appropriate boundary conditions, that amount to 
choosing $G(x;x')=G_{\rm ret}(x;x')$ and fixing once and for all
the homogeneous solution. In particular, one should be careful not to use the corresponding Lagrangian at the quantum level, and one should not include the corresponding fields in the external lines or in the loops. The corresponding particles, some of which are unavoidably ghost-like, do not correspond to propagating degrees of freedom in our original problem, and the quantization of these spurious solutions does not make sense.

We have seen in particular how the above considerations apply to the model
defined by \eq{modela2}. We have found that the apparent ghost signaled by the second term in \eq{TT} is actually a non-radiative degree of freedom, and we have also seen that in the $m\ra 0$ limit it goes smoothly into a non-radiative degree of freedom of GR. Finally,  we have shown how the same conclusion emerges from the point of view of the spurious degrees of freedom induced by the localization procedure. The conclusion is that the apparent ghost of \eq{TT} is not an indication of  any problem of consistency of the theory at the quantum level, and \eq{modela2}, taken as an effective classical equation, defines a consistent classical theory that can be safely used for cosmological purposes.

\vspace{5mm}

\noindent
{\bf Acknowledgments.} We thank  Claudia de~Rham, Stanley Deser, Yves Dirian,
Lavinia Heisenberg and Maud Jaccard  for useful discussions. Our work is supported by the Fonds National Suisse.

\bibliographystyle{utphys}
\bibliography{myrefs_massive}

\end{document}